\documentclass[twocolumn,twocolappendix]{aastex631}

\usepackage{amsmath}
\usepackage{txfonts}
\usepackage{xcolor}
\usepackage{graphicx}
\usepackage{gensymb}
\usepackage{booktabs}
\usepackage{bm}
\usepackage{lipsum}
\usepackage{float}

\newcommand{\jwst}[1]{%
  \emph{JWST}\
}

\newcommand{\hst}[1]{%
  \emph{HST} %
}

\newcommand{\citepap}[1]{%
  {\color{orange} [CITE]} %
}

\begin{document}

\title{Efficient survey design for finding high-redshift galaxies with JWST}

\author[0000-0001-7697-8361]{Luka Vujeva}
\affiliation{Niels Bohr International Academy, Niels Bohr Institute, Blegdamsvej 17, DK-2100 Copenhagen, Denmark}
\affiliation{Cosmic Dawn Center (DAWN)}
\affiliation{Niels Bohr Institute, University of Copenhagen, Jagtvej 128, DK-2200, Copenhagen N, Denmark}

\author[0000-0003-3780-6801]{Charles L. Steinhardt}
\affiliation{Cosmic Dawn Center (DAWN)}
\affiliation{Niels Bohr Institute, University of Copenhagen, Jagtvej 128, DK-2200, Copenhagen N, Denmark}

\author[0000-0002-8896-6496]{Christian Kragh Jespersen}
\affiliation{Department of Astrophysical Sciences, Princeton University, Princeton, NJ 08544, USA}

\author[0000-0003-1625-8009]{Brenda L.~Frye}
\affiliation{Department of Astronomy/Steward Observatory, University of Arizona, 933 N. Cherry Avenue, Tucson, AZ 85721, USA}

\author[0000-0002-6610-2048]{Anton M. Koekemoer}
\affiliation{Space Telescope Science Institute, 3700 San Martin Dr., Baltimore, MD 21218, USA}

\author[0000-0002-7809-0881]{Priyamvada Natarajan}
\affiliation{Department of Astronomy, Yale University, New Haven, CT 06511, USA}
\affiliation{Department of Physics, Yale University, New Haven, CT 06520, USA}

\author[0000-0002-9382-9832]{Andreas L. Faisst}
\affiliation{Caltech/IPAC, 1200 E. California Blvd. Pasadena, CA 91125, USA}

\author[0000-0002-9120-234X]{Pascale Hibon}
\affiliation{European Southern Observatory, Alonso de Cordova 3107, Vitacura, Santiago, Chile}

\author[0000-0001-6278-032X]{Lukas J. Furtak}
\affiliation{Physics Department, Ben-Gurion University of the Negev, P.O. Box 653, Be'er-Sheva 84105, Israel}

\author[0000-0002-7570-0824]{Hakim Atek}
\affiliation{Institut d'Astrophysique de Paris, CNRS, Sorbonne Universit\'e, 98bis Boulevard Arago, 75014, Paris, France}

\author[0000-0002-7570-0824]{Renyue Cen}
\affiliation{Institute for Advanced Study in Physics, Zhejiang University, Hangzhou 310027, China and Institute of Astronomy, Zhejiang University, Hangzhou 310027, China}

\author[0000-0002-5460-6126]{Albert Sneppen}
\affiliation{Cosmic Dawn Center (DAWN)}
\affiliation{Niels Bohr Institute, University of Copenhagen, Jagtvej 128, DK-2200, Copenhagen N, Denmark}






\begin{abstract}

%
%
Several large \emph{JWST} blank field observing programs have not yet  discovered the first galaxies expected to form at  $15 \leq z \leq 20$. This has motivated the search for more effective survey strategies that will be able to effectively probe this redshift range.
%
%
Here, we explore the use of gravitationally lensed cluster fields, that have historically been the most effective discovery tool with \emph{HST}. In this paper, we analyze the effectiveness of the most massive galaxy clusters that provide the highest median magnification factor within a single \jwst\ NIRCam module in uncovering this population.
%
%
The results of exploiting these lensing clusters to break the $z > 15$ barrier are compared against the results from large area, blank field surveys such as JADES and CEERS in order to determine the most effective survey strategy for \emph{JWST}.
%
%
We report that the fields containing massive foreground galaxy clusters specifically chosen to occupy the largest fraction of a single NIRCam module with high magnification factors in the source plane, whilst containing all multiple images in the image plane within a single module provide the highest probability of both probing the $15 \leq z \leq 20$ regime, as well as discovering the highest redshift galaxy possible with \emph{JWST}. 
We also find that using multiple massive clusters in exchange for shallower survey depths is a more time efficient method of probing the $z>15$ regime. 

\end{abstract}



\section{Introduction} \label{sec:intro}

\jwst\ was designed to have the capacity to observe and detect the very first generation of galaxies during their process of assembly, rather than in their evolved state. Although we do not have direct observational constraints on the assembly of the first galaxies at present, the formation of halos that likely host those galaxies is robustly predicted both analytically and numerically. Theoretical models in the context of the standard paradigm for structure formation in the dark energy - dark matter dominated Universe, predict that the first galaxies should only appear between redshifts of $15 \leq z \leq 20$ \citep{Springel_2005,Bromm_2011,Lacey_2011,2023arXiv230404348Y}. Pinpointing the timing of this first generation will provide a stringent test of different halo assembly scenarios and place further constraints on the age of the Universe at which the first stars formed. Additionally, it will allow us to pinpoint mark the beginning of the cosmic ``dark ages'', in which neutral hydrogen obscures the light from the first galaxies, thus preventing us from directly observing them \citep{Mason_2015}.

However, finding these galaxies requires an optimized search strategy. The early success of \jwst\ has allowed us to observe a new regime of high-redshift galaxies in both blank and cluster lens fields, which have been spectroscopically confirmed to lie at redshifts of up to $z = 13.2$ \citep{robertson2022discovery}. 

Cluster fields have also been successful in finding a plethora of new robust high-redshift candidates up to redshifts of $z\sim12$ \citep{Atek_2022, pascale2022, Adams_2022, noirot2023} in surveys such as CANUCS (GTO $\#1208,2779,4527$, PI Willot) \citep{2017jwst.prop.1208W}, PEARLS (GTO $\#1176,2738$, PI Windhorst) \citep{Windhorst_2022}, UNCOVER (GO $\#2561$, PI Labbe) \citep{uncover2022}, GLASS-JWST-ERS (ERS $\#1324$, PI Treu) \citep{Treu_2022,paris2023}, and ERO observations of the cluster SMACS 0723 \citep{Pontoppidan_2022}.
They have even been successful in identifying an overdensity of bright galaxies at $z\sim10$ magnified by Abell 2744 \citep{weaver2023, castellano2023}, as well as even identifying the first lensed red giant star at cosmological distances along the line of sight of ACT-CL J0102-4915, otherwise known as "El Gordo" \citep{Diego2023,frye2023}. This is rapidly approaching the predicted redshift window for the birth of the first galaxies in the Universe, however, we are yet to directly observe the predicted beginning of the dark ages. 
%

%
%
In the blank fields in which the highest redshift galaxies to date have been found \citep{robertson2022discovery}, the maximum redshift at which a galaxy is statistically likely to be found is $z \sim 14$ \citep{Steinhardt_2021}.
However, scarce attention has been paid to optimize search strategies to glimpse into the target $15 \leq z \leq 20$ regime which \jwst\ should be capable of probing.
This is a critical difference because various models predict that the redshift at which we can first probe the cosmic dark ages lies between $z \sim 15$ and $ z \sim 20$.
In this paper, we present a new method  that uses galaxy clusters that provide the highest median magnification factor within a single NIRCam module in order to give us the highest probability of finding a galaxy within the accessible redshift range.

A major hindrance to finding high-redshift galaxies is cosmic variance.
Cosmic variance increases strongly towards both high mass and high redshift \citep{2001MNRAS.323....1S, Moster_2011,Bhowmick_2020,Steinhardt_2021}.  Primordial fluctuations produce a small number of extremely overdense regions which will lead to the quickest gravitational collapse and the first, massive galaxies. These extreme overdensities are tightly clustered, so that most of the first galaxies will appear in a small number of rich fields\footnote{This is analogous to searching for the tallest mountains on the surface of the Earth, since any very tall mountain is likely to exist along a general "overdensity" of mountains in a range, and one is thus disproportionately more likely to find several tall mountains if one is found.}.  That is, the relatively uncommon pointings which have more high-mass, ultra-high redshift galaxies than average are overwhelmingly likely to have more galaxies any mass at the same redshifts as well.  Conversely, the regions containing lower-mass galaxies must still have extreme overdensities in order for their halos to have collapsed so quickly, and thus are very likely to have high-mass galaxies nearby.

Thus, although finding these regions is potentially difficult and may require probing many different sightlines, the discovery of even a single, massive ultra-high redshift galaxy is likely to indicate the presence of many smaller galaxies nearby.

%
In this work we revisit the abilities of a well known class of cluster lenses to augment the maximum survey depth of \emph{JWST}, and examine how to best exploit them as tools to discover the first galaxies. The paper is structured as follows: Section~\ref{sec:modelandmethod} describes the models required to carry out this work, as well as the methods in which they are implemented into our testing, Section~\ref{sec:strategy} outlines the two different survey types that are being compared, as well as our criterion for selecting the ideal survey, and Section~\ref{sec:idealstrategy} summarizes our findings, as well as explores future work.

This work adopts a standard flat $\Lambda$CDM cosmological model with $\Omega_m = 0.3$ and $h = 0.7$, as well as an AB magnitude system \citep{oke1974, gunn1983}. 

\section{Models And Methodology}\label{sec:modelandmethod}

This work employs three key models in order to achieve its end result: 1) the extrapolated luminosity function into the $z>8$ regime, and 2) the lens models of massive galaxy clusters modelled using \emph{HST}/\emph{JWST} data. This section will explore these models and the methods in which they were used to build our mock survey code.

\subsection{The Luminosity Function}\label{subsec:lumfunction}

The luminosity function (LF), which is the number density of galaxies $\phi(M)$ as a function of their apparent UV luminosity ($m_{UV}$) was extrapolated from high-redshift \emph{HST} surveys to estimate the $z>8$ regime. \citep{Bouwens_2015} found that the resulting luminosity function after fitting observed luminosity functions for $z \sim 4-10$ with a Schechter function were specified as follows\citep{schechter76}:
 \begin{equation}
 \begin{split}
    \phi (M) & = \phi^* \frac{ln(10)}{2.5}10^{-0.4(M - M^*)(\alpha + 1)}e^{-10^{-0.4(M - M^*)}} \\
    M^*_{UV} & = (-20.95 \pm 0.10) + (0.01 \pm 0.06)\times(z - 6) \\
    \alpha & = (-1.87 \pm 0.05) + (-0.10 \pm 0.03)\times(z - 6) \\
    \phi^* & = (0.47^{+0.11}_{-0.10})\times10^{(-0.27 \pm 0.5)(z-6)}10^{-3} {\rm Mpc^{-3}} \ .
\end{split} 
\end{equation}

These fits were used to extrapolate the LF to $z>10$, and the substantial uncertainties at $z \gg 6$ were reported in Fig. 3 of \cite{Steinhardt_2021}. The UV LF was converted to a stellar-mass LF in accordance with \cite{Song_2016}, which fit a linear relation between $M_{UV}$ and log($M_* / M_\odot$). Following \cite{Song_2016}, the best fit slope at $z=8$ was used for $z>8$ extrapolations, along with the intercepts of the linear fit of the data between $4<z<8$, resulting in a universal mass-to-light ratio for a fixed redshift.


\subsection{Selecting the optimal foreground clusters for finding ultra-high redshift galaxies} \label{sec:clusterparams}

Since the background galaxies being magnified lie at a greater distance, the optimal foreground clusters for lensing the $z \sim 15-20$ galaxies we hope to discover with \emph{JWST} will be different from those optimal for finding $z \lesssim 10$ galaxies with \emph{Hubble}. To select clusters, we consider the most massive potential foreground clusters, estimating magnification maps. Then, extrapolated luminosity functions and an updated cosmic variance calculator \citep{Moster_2011, Steinhardt_2021} were used to calculate the probability distribution for discovering highest-redshift galaxies using these magnification maps. The calculator is empirically calibrated with a mass-dependent scaling to the dark matter cosmic variance. The calibration assumes that stellar masses and halo masses are related via abundance matching which introduces an uncertainty of $\approx0.3$ dex in the mass bins \citep{Jespersen2022_mangrove, Chuang2023_AM}.

\begin{figure}[ht!]
    \begin{center}
        \includegraphics[width= \linewidth]{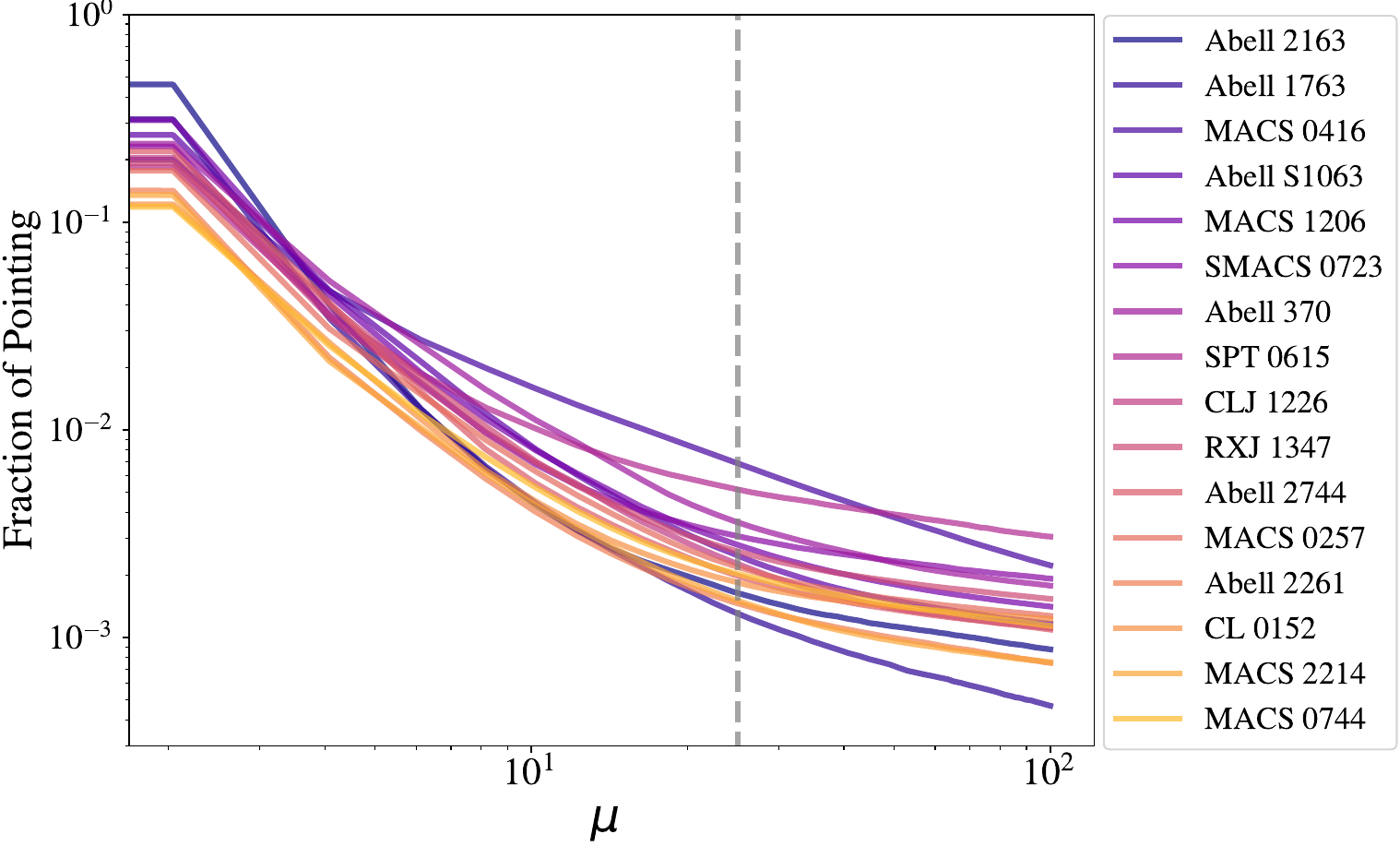}
        \caption{Fraction of NIRCam pointing covered by magnification factors greater than $\mu$ for a source at $z=15$ (in the source plane). 
        The vertical gray line corresponds to the $\mu=25$ threshold that was placed on the magnification maps in order to mitigate the errors associated with the high-magnification regions of the lens models. The clusters that were selected occupy the highest fraction of pixels within a module with high magnification factors.
        }
        \label{fig:permu}
    \end{center}
\end{figure}

Here, we select the foreground clusters based on two simple metrics: the fraction of pixels within a single 2.2' $\times$ 2.2' module \footnote{https://jwst-docs.stsci.edu/jwst-near-infrared-camera} that fall above a given magnification factor (Fig.~\ref{fig:permu}), and the probability distribution of the highest redshift galaxy discovered, optimized for $15 \leq z \leq 20$, the median highest redshift found in simulated lensed pointings. This generally equates to the most massive, and most concentrated galaxy clusters \citep{Comerford_2007, Merten_2015}. Thus, the clusters in this work provide the highest probability of magnifying a background source due to them covering the highest fraction of the source plane with large magnification factors (and in turn high effective survey volumes \citep{Atek_2018}).  
 
\begin{deluxetable*}{lccccc}[ht]
\tablecaption{Summary of clusters selected from Hubble Frontier Field, CLASH and RELICS programs selected for this study, arranged in order of increasing redshift, along with the corresponding reference for each lens model used in this work.
}
\label{table:clusterlist}
\tablewidth{0pt}
\tablehead{
\colhead{Cluster Name} & \colhead{Redshift ($z$)} & \colhead{RA (deg/J2000)}& \colhead{Dec. (deg/J2000)} & \colhead{Model Used in this Work} }
\startdata
Abell 2163  & $0.203$ &  $243.9541667$  & $-6.1522222$ &  \citet{cerny2018} \\
Abell 2261  & $0.225$ & $ 260.61336$ & $32.232465$ &  \citet{Zitrin_2015} \\
Abell 1763  & $0.228$ & $203.8000000$ & $41.0002778$ &  \citet{salmon2020}  \\  
Abell 2744  & $0.308$ & $3.5862583$ & $-30.4001750$  &  \citet{Richard_2021}\\
MACS J0257.6-2209   & $0.322$ & $44.4211250$ & $-22.1549167$ &  \citet{Richard_2021} \\
Abell S1063  & $0.351$ & $342.2262500 $ & $-44.5186111$ & \citet{Zitrin_2015} \\
Abell 370  & $0.375$ & $39.9713417$ & $-1.5822611$ & \citet{Richard_2021} \\ 
SMACS J0723.3-7327  & $0.390$ & $110.8054167$ & $-73.4569444$ & \citet{salmon2020} \\
MACS J0416.1-2403  & $0.397$ & $64.0381000$ & $-24.0674861$ & \citet{Richard_2021}  \\
MACS J1206.2-0847  & $0.438$ & $181.55065$ & $-8.8009395$ &  \citet{Richard_2021} \\
RX J1347.5-1145   & $0.451$ & $ 206.87756$ & $-11.752610$  &  \citet{Richard_2021}\\
MACS J2214.9-1359  & $0.502$ & $333.7387167 $ & $-14.0035861 $  & \citet{Richard_2021} \\
MACS J0744.9+3927  & $0.686$ & $116.22000$ & $39.457408$  & \citet{Zitrin_2015} \\
CL 0152-13  & $0.833$ & $28.1666667$ & $-13.9552778$ &  \citet{Acebron_2019} \\
CL J1226.9+3332  & $0.890$ & $186.74270$ & $33.546834$ &  \citet{Zitrin_2015}\\
SPT 0615-57& $0.972$ & $93.9833333 $ & $-57.7638889$ &  \citet{mahler_2018} \\
\enddata   
\end{deluxetable*}

%
%
%


The candidate clusters were selected from the Hubble Frontier Fields \citep{2017HFF}, RELICS \citep{cerny2018,Coe_2019,salmon2020}, and  CLASH \citep{Postman_2012} program and have VLT/MUSE observations \citep{Richard_2021}. 
Clusters were selected from these surveys due to their containing the most massive known galaxy clusters, and their plethora of multi-wavelength observations, which has enabled accurate lens modelling.
%
%
%
%


More generally, these clusters will magnify the highest fraction of their pointing for any large magnification factor $\mu$ (Fig.~\ref{fig:permu}), making them the most efficient foreground clusters for finding targets requiring higher magnification, and thus the best new discovery tool for fainter, ultra-high redshift sources as well.

An additional effect of the high cosmic variance is that a small fraction of pointings will contain a strong overabundance of massive, ultra-high redshift galaxies, while most of the remainder will contain none.  Although the disadvantage is that this requires many independent sightlines for an initial discovery, it also implies that a single, massive ultra-high redshift galaxy is likely to lie in a rich field with many lower-mass companions at similar redshift.  Thus, a robust discovery of a single high-mass, ultra-high redshift galaxy can be used as a signpost indicating the way to an ideal target for deeper followup observations.  

It should be noted that this strategy will necessarily discover galaxies only in extremely overdense and thus atypical regions. However, at present, there is no effective search strategy to find lower mass $z>15$ galaxies in typical regions, since blank field ultra deep surveys are unlikely to find them. Thus, a biased survey which will require significant modelling and correction is currently the only option.

\subsection{Galaxy Cluster Models}\label{subsec:clustermodel}

The galaxy cluster models were generated from the publicly available products from the CLASH \citep{2015ApJ...801...44Z, Richard_2021} (some with followup MUSE observations \citep{Richard_2021}), and RELICS \citep{cerny2018, salmon2020} programs, which utilized combinations of strong and weak gravitational lens modelling codes such as \verb|LENSTOOL|, which utilizes the LTM method outlined in \cite{Zitrin_2009,Zitrin_2015,Broadhurst_2005, 2015ApJ...801...44Z}, as well as \textit{glafic} \citep{glafic2010,glafic2019}. The light-traces-mass assumption (LTM) assumes that the mass distribution of both the cluster and member galaxies is reasonably traced by the cluster's light distribution \citep{Broadhurst_2005, 2015ApJ...801...44Z}, and is used as a starting point for fitting the locations and distributions of dark matter in the cluster. 
%
%

\begin{figure*}
    \begin{center}
        \includegraphics[width=\linewidth]{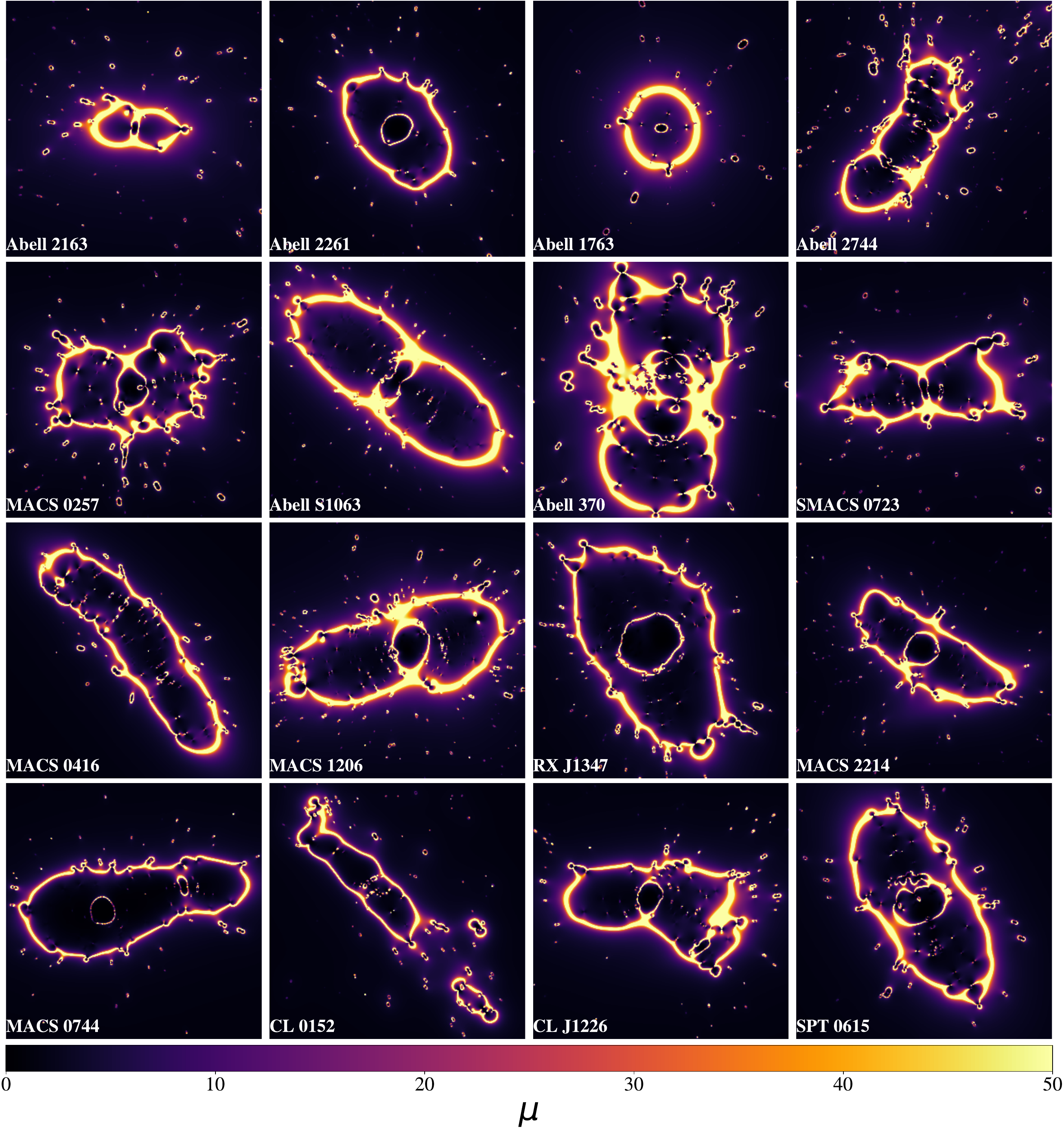}
        \caption{
        Magnification maps ($\mu$) of the foreground clusters used in this work as seen in the image plane of a single NIRCam module (2.2'x2.2'), scaled for a source at $z=15$, and arranged in ascending redshift.
        }
        \label{fig:mumapsimage}
    \end{center}
\end{figure*}

\begin{figure*}
    \begin{center}
        \includegraphics[width=\linewidth]{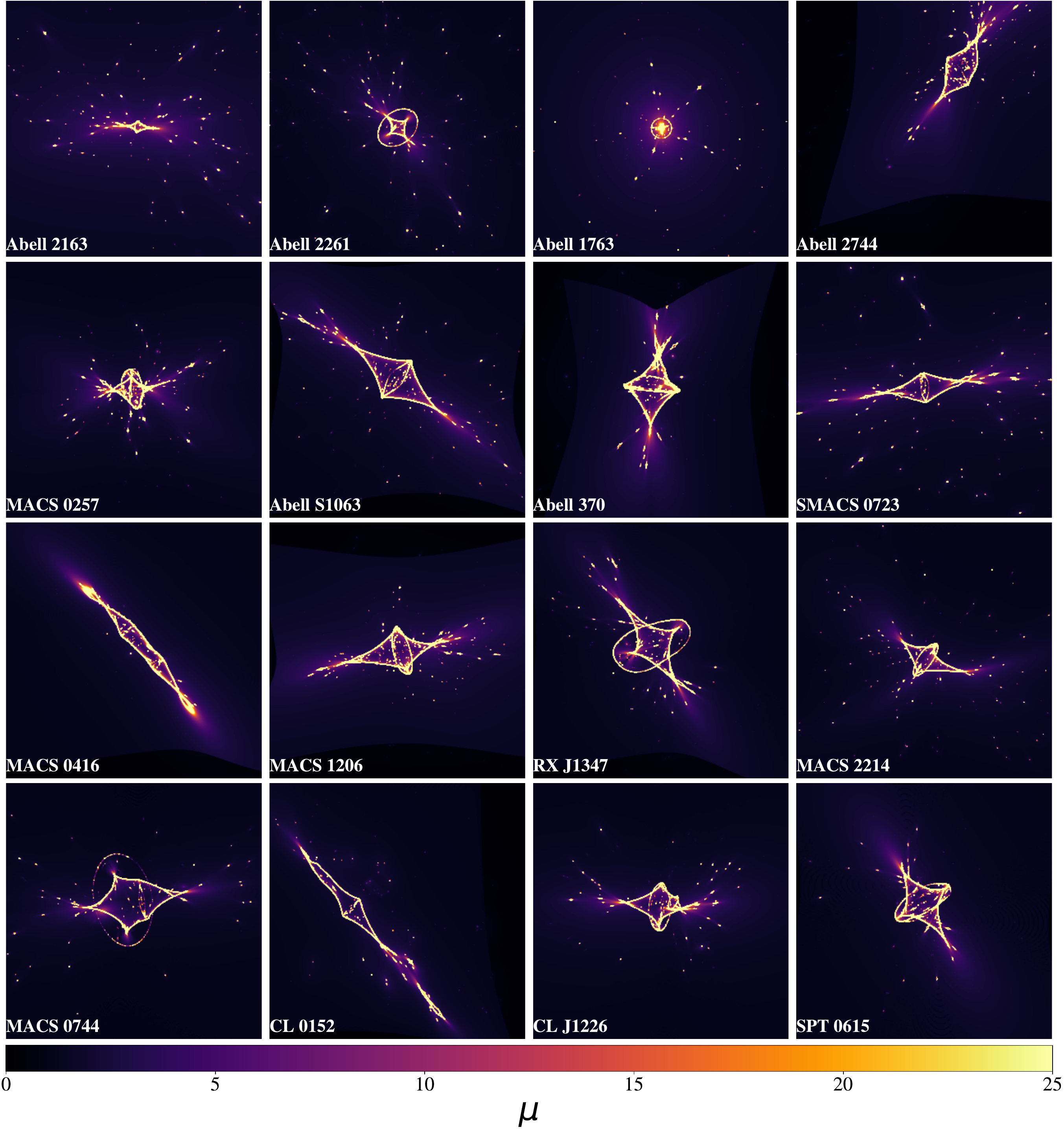}
        \caption{
        Like Fig.~\ref{fig:mumapsimage}, but in the source plane.
        }
        \label{fig:mumaps}
    \end{center}
\end{figure*}

%
The lens models used in this work are shown in both the image plane and the source plane in  Fig.~\ref{fig:mumapsimage} and Fig.~\ref{fig:mumaps} respectively. The resulting magnification maps were then re-scaled for a source at $z=15$ (due to there being a lack of significant differences between the magnification maps of sources at $z > 8$), limited to a maximum magnification factor of $\mu=25$ in order to limit the significant errors associated with the extremely high magnification regions in the models, and cut to the size of a single \emph{JWST} NIRCam module of 2.2'x2.2'.



\subsection{Implementation of Magnification Maps}\label{subsec:mumapapplication}

The magnification maps corresponding to either the galaxy cluster or elliptical galaxy lens models were implemented into our mock survey code in order to model the benefits that their respective magnification effects have in detecting faint high-redshift sources. Although \citet{Steinhardt_2021} took the mean magnification factor of a lensed field into account in their mock survey trials, they did not take into account the randomness that is inherently associated with searching for high-redshift targets behind foreground lenses. As seen in Fig.~\ref{fig:mumaps}, the magnification factors vary drastically within the field of view, therefore the mean or median magnification factor is not an adequate representation of the effects of the magnification factors within the field.

The mock survey code used in \citet{Steinhardt_2021} computed the mean number of objects one would expect to see within a given survey volume (before taking cosmic variance into account) by integrating the luminosity function for every mass bin over the range of apparent magnitudes that would be visible by the given instrument. In this case, the maximum limiting magnitude of JWST $m=31.5$ was used. An important note is that the mass bins were chosen to be the same as those used by \citet{Moster_2011} in order to be able to calculate the cosmic variance associated with those masses ($M = [7.0-11.0] \  \log{(M_\odot)}, \ \Delta M = 0.5 \  \log{(M_\odot)}$). This work also made use of the \verb|cosmic-variance|\footnote{https://pypi.org/project/cosmic-variance/} python package, which is a python adaptation of the original IDL code released with \citet{Moster_2011}.

In order to attempt to model the random locations of background sources, during the step where the luminosity function is integrated up until a fixed limiting magnitude, a random pixel from the magnification map is drawn, and the mean magnification factor within an area corresponding to the expected size of a source  \citep[R$_{\rm eff} \sim 0.5$ kpc;][]{adamssmacsdiscovery} was taken, and converted into the new limiting magnitude

\begin{equation}
    m_{\rm lim, magnified} = m_{\rm lim,intrinsic} + 2.5\log{(\mu)} \ ,
\end{equation}

where $m_{\rm lim, intrinsic}$ is the limiting magnitude of the survey and $\mu$ is the random magnification factor drawn from the magnification map of the foreground cluster. The process was repeated over all redshift bins in order to introduce as many independent source locations within the pointing as possible. 
%
%


\section{Comparative method for selecting the ideal survey strategy}\label{sec:strategy}

Two survey types were compared in order to determine their relative efficiencies in finding high-redshift galaxies: using foreground massive galaxy clusters, and a larger "snapshot" style of survey with a lower integration time but significantly larger search area, such as the JADES Deep and Medium surveys \citep{robertson2022discovery}, which have proven to be successful in finding $z \sim 13$ galaxy candidates using a combination of photometry and spectroscopy.
The clear benefit that the snapshot program has to its lensed counterpart is the increase in survey area and volume, which helps combat the effects of cosmic variance at high-redshifts.


Testing was conducted using a modified version of the code created by \citet{Steinhardt_2021}, which was updated to accept magnification maps from various foreground lenses in order to simulate the magnification effects from strong gravitational lenses on the number of galaxies found, and the maximum distance redshift ($z_{max}$) we can expect to see galaxies in each field. The four main results which were used to compare the efficacy of each survey design were: the distribution of magnification factors above a given threshold within the field; the median magnification factor of each cluster within the single pointing; the mean and median highest redshift galaxy found in each iteration ($z_{max}$); and the number of galaxies found within each redshift bin.


As per the findings of \citet{Steinhardt_2021}, proposed JWST surveys of blank fields, such as JADES or CEERS, are likely to find galaxies up to a median redshift of $z \sim 12-14$ depending on the survey strategy.

Early releases from the JADES survey have already yielded a record breaking galaxy JADES-GS-z13-0 at $z_{spec} = 13.2$ \citep{robertson2022discovery}, which was originally discovered photometrically using NIRCam, but whose Lyman break was confirmed spectroscopically using NIRSpec shortly afterwards. 


Both of these early \emph{JWST} surveys have proven to be successful in finding $9\leq z \leq 13$ galaxies, in part because they employ wider search areas (46--190 arcmin$^2$), with limiting magnitudes of 30.7--29.8 mag, respectively. The benefits of such a survey strategy is that the effects of cosmic variance are slightly weaker with large survey volumes. However, both of these surveys required large time allocations in order to achieve high limiting magnitudes over large survey areas.
Additionally, the bi-modality prevalent in the photometric estimates of ultra-high redshift galaxies \citep{Coe_2019,Treu_2022,steinhardt2023highestredshift, haro2023spectroscopic} will require follow-up spectroscopy to robustly calculate the redshift of the galaxies. The necessity for high-quality follow-up spectroscopy will be greatly aided by observing lensed fields, due to the magnification from the foreground lens increasing the apparent brightness and size of the background source, thus reducing the integration time required to resolve it.
Because cosmic variance favors independent lines of sight as opposed to large spatially coherent search areas, the search for high-z galaxies would have greatly benefited from independent lines of sight with similar time commitments \citep{Trenti_2011}.


\section{Conclusions}\label{sec:idealstrategy}

Our study compares the results of two survey strategies: blank field snapshot, and galaxy cluster lens survey strategies, and concludes that the clear choice for a survey optimized for finding ultra-high redshift galaxies utilizes multiple massive foreground galaxy cluster lenses. 


\begin{figure}[h!]
    \begin{center}
        \includegraphics[width=\linewidth]{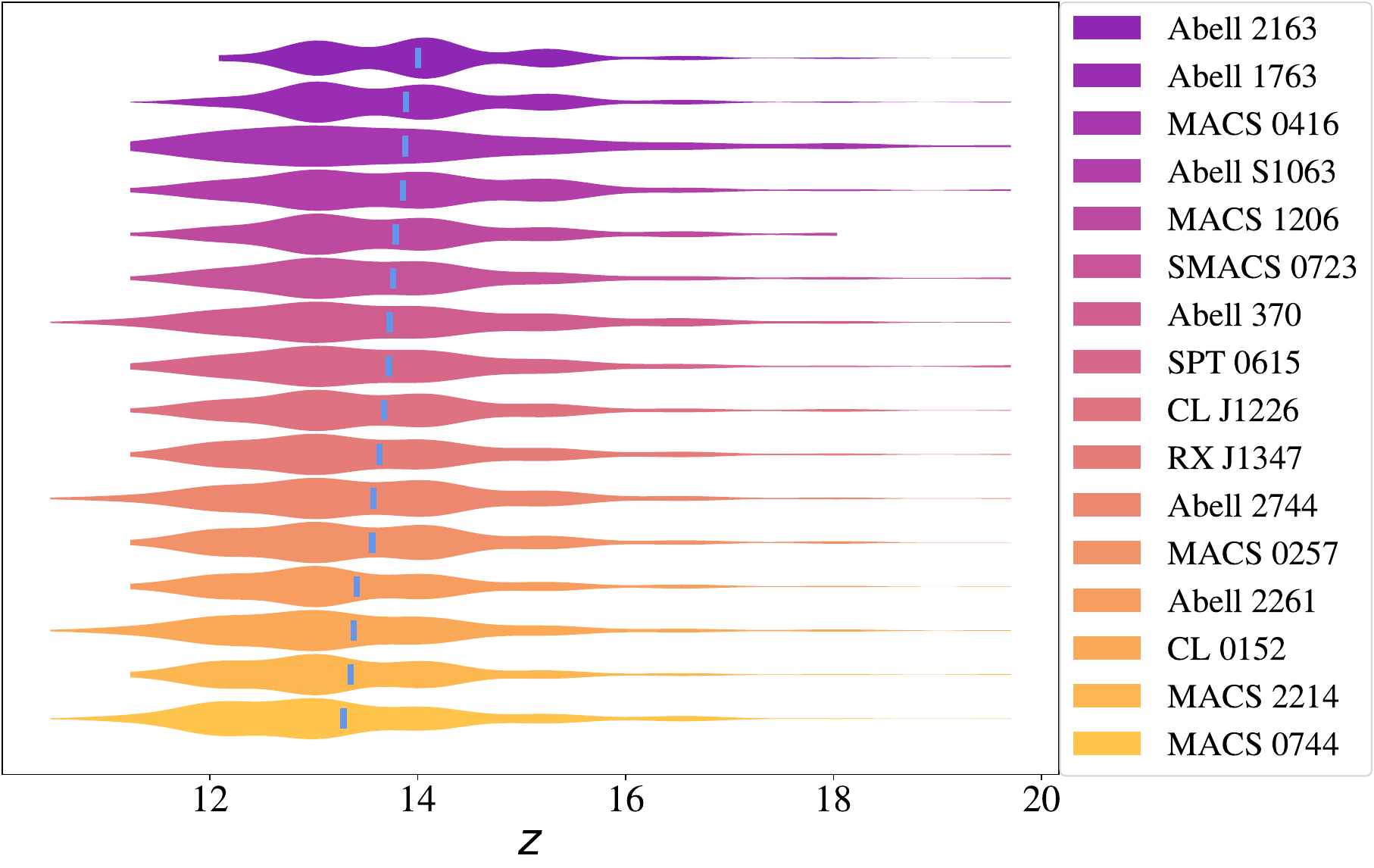}
        \caption{Distribution of the highest redshift galaxies found in  500 random realisations of each cluster at a depth of $m_{lim} = 31.5$. The vertical blue lines represent the mean highest redshift galaxy found along each cluster line of sight. Note that although the means lie just below $z=14$, the probability distributions of all of the cluster lines of sight extend past $z>15$.
        }
        \label{fig:zmax}
    \end{center}
\end{figure}

%

\begin{deluxetable}{lcccc}
\tablecaption{Median magnification factor in the source plane, mean highest redshift galaxy found in the simulated survey for each line of sight along each foreground cluster, as well as the probability of finding a galaxy above $z \sim 14$ and $15$ over 500 trials.
}
\label{table:clusterresults}
\tablewidth{0pt}
\tablehead{
\colhead{Cluster Name} &\colhead{ $\mu_{\textrm{med}}$} & \colhead{{$\big \langle z_{max} \big \rangle $}}& \colhead{$P(z > 14)$}& \colhead{$P(z > 15)$}  }
\startdata
Abell 2163          & $1.47$ & $13.41$ & 62.0\% & 14.0\% \\
Abell 2261          & $1.99$ & $14.00$ & 37.0\% & 23.2\% \\
Abell 1763          & $1.82$ & $13.89$ & 54.0\% & 21.6\% \\
Abell 2744          & $1.61$ & $13.57$ & 46.0\% & 18.0\% \\
MACS J0257.6-2209   & $1.52$ & $13.56$ & 44.6\% & 16.6\% \\
Abell S1063         & $1.60$ & $13.86$ & 53.0\% & 24.8\% \\
Abell 370           & $1.51$ & $13.73$ & 47.4\% & 23.2\% \\ 
SMACS J0723.3-7327  & $1.59$ & $13.76$ & 47.6\% & 18.4\% \\
MACS J0416.1-2403   & $1.43$ & $13.88$ & 48.8\% & 24.8\% \\
MACS J1206.2-0847   & $1.75$ & $13.79$ & 49.4\% & 20.6\% \\
RX J1347.5-1145     & $1.49$ & $13.63$ & 43.6\% & 18.6\% \\
MACS J2214.9-1359   & $1.44$ & $13.35$ & 36.2\% & 11.8\% \\
MACS J0744.9+3927   & $1.33$ & $13.29$ & 33.6\% & 13.8\% \\
CL 0152-13          & $1.33$ & $13.39$ & 37.2\% & 14.4\% \\
CL J1226.9+3332     & $1.57$ & $13.67$ & 46.6\% & 18.0\% \\
SPT 0615-57         & $1.53$ & $13.72$ & 46.6\% & 19.8\% \\
\enddata   
\end{deluxetable}

\begin{deluxetable}{lccccccc}
    \tablecaption{Highest redshift galaxy found in the median and mean simulated blank field surveys, as well as the percentage of finding a galaxy above $z \sim 14$ over 500 trials. Note that the mean and median $z_{max}$, as well as the probability of finding an ultra-high redshift galaxy are significantly lower than those of the optimal cluster fields in Tab.~\ref{table:clusterresults}. $P(z > 15)$ is negligible in these surveys.
    }
    \label{table:zmaxblankabove14}
    \tablewidth{0pt}
    \tablehead{
    \colhead{Survey} & \colhead{Median {$z_{max}$}} & \colhead{{$\left<z_{max}\right>$}}  & \colhead{$P(z > 14)$} }
    \startdata
    JADES (M)  & $12.09$ & $12.27$ & $1.8\%$   \\
    JADES (D)  & $13.03$ & $12.95$ & $17.2\%$  \\ 
    CEERS  & $10.47$ & $10.48$ & $0\%$   \\
    \enddata   
\end{deluxetable}

As summarized in Tables~\ref{table:clusterresults} and~\ref{table:zmaxblankabove14}, lines of sight containing galaxy clusters optimized for magnifying high-redshift galaxies significantly outperform \emph{JWST} blank field surveys such as JADES or CEERS, which are specifically designed to find high-redshift galaxies.

The pointings containing the ideal galaxy clusters not only provide the highest probability of finding an ultra-high redshift galaxy, but the strong magnification factors present along these lines of sight could allow for the highest redshift galaxy that \emph{JWST} could possibly detect. This is in spite of the strong effects of cosmic variance in the $15 \leq z \leq 20$ regime, which are lower in the large area blank field surveys. 

Another distinct benefit to using single, deep JWST pointings as opposed to combining many of them side by side in order to cover large patches of the sky is that one could observe a multitude of clusters in the same time that it takes to observe one large patch of the sky.

For example, if it takes 10 hours of NIRCam photometry to reach a given limiting magnitude in a single pointing, a blank field survey comprised of 10 NIRCam pointings such as CEERS would require 100 hours of photometry (not including overheads) in order to detect a galaxy up to $z\sim14$. Even in this case, the pointings are side by side and not independent, therefore not negating the effects of cosmic variance in the most optimal manner for the high-redshift regime. However, if you were to use the same 100 hours of photometry time looking at galaxy clusters, you would be able to observe the 10 best clusters (which can be seen in Fig.~\ref{fig:zmax}) at the same depth in the same amount of time. Not only are you statistically more likely to find a $z\geq 15$ galaxy in this manner (shown in Table~\ref{table:clusterresults}), but you would benefit from completely independent lines of sight between the different clusters, thus providing the added benefits of optimally combating cosmic variance.

\subsection{Shallow depths in exchange for more lines of sight}

It is important to consider that even after the large integration times required to reach NIRcam's maximum limiting magnitude, one might still not see any high redshift galaxies along that line of sight. Therefore, the most efficient survey strategy for \emph{JWST} might not be to simply observe the single best galaxy clusters at heroic depths, but instead to survey a multitude of the best clusters at slightly shallower depth.
\citet{salmon2020} showed  that medium depth observations of a large number of galaxy clusters was a successful strategy in obtaining large amounts of photometric $z>5.5$ candidates with \emph{HST}. 
This is most likely due to a combination of the sharpness of the caustics in the source plane (making it less likely for a background galaxy to lie on a caustic along a single line of sight), coupled with the strong effects of cosmic variance at high redshifts, which will be even stronger with \emph{JWST} probing a higher redshift regime than \emph{HST}. 

 Even in the most ideal cases where the magnification factors in the source plane (Fig.~\ref{fig:mumapsimage},\ref{fig:mumaps}) cover as much of a NIRCam module as possible (whilst still being able to contain all of the multiple images within the pointing in the image plane), the probability of a background galaxy lying in the strong lensing regime is very low. Therefore, by increasing the number of independent lines of sight by looking at a variety of clusters, you increase the chances of a high-redshift galaxy lying along one of the caustics.
 Given the strength of the magnification factors along the caustics (Fig.~\ref{fig:mumaps}), even at lower limiting magnitudes, one could still get large enough magnification factors to observe high-redshift galaxies that would otherwise be un observable in blank fields. 

In order to compare the effects of depth and independent lines of sight with respect to the probability of finding a $z>15$ galaxy, we use the magnification map of SMACS 0723 (due to it representing the average of all of the clusters in this study in terms of lensing efficiency), and repeat the mock survey at incrementally decreasing steps in limiting magnitude until there is almost no chance of finding a $z>15$ galaxy. 
%
%
We use the \emph{JWST} exposure time calculator \citep{2016SPIE.9910E..16P} (ETC v3.0) \footnote{\url{https://jwst-docs.stsci.edu/jwst-exposure-time-calculator-overview}} to estimate the exposure time required to obtain roughly a $5\sigma$ observation at a variety of depths with the 6 NIRcam filters already used to observe SMACS 0723 with \emph{JWST} \citep{Pontoppidan_2022}: F090W, F150W, F200W, F277W, F356W, and F444W. Although a combination of photometric and spectroscopic observations is the gold standard for calculating redshifts of galaxies (especially at high redshifts), photometry can still provide good estimates \citep{steinhardt2022templates}, and can help identify high-z candidates which can be confirmed with follow-up spectroscopy. 
Finally, we multiply the exposure time for each depth by the number of independent lines of sight to compare the total exposure time required to achieve a given probability of finding a $z>15$ galaxy in the cluster field. Due to there being a limited number of additional cluster lenses that match the efficiency of those presented in this work, the maximum number of clusters lines of sight was limited to 25.


\begin{figure}[h!]
    \begin{center}
        \includegraphics[width=\linewidth]{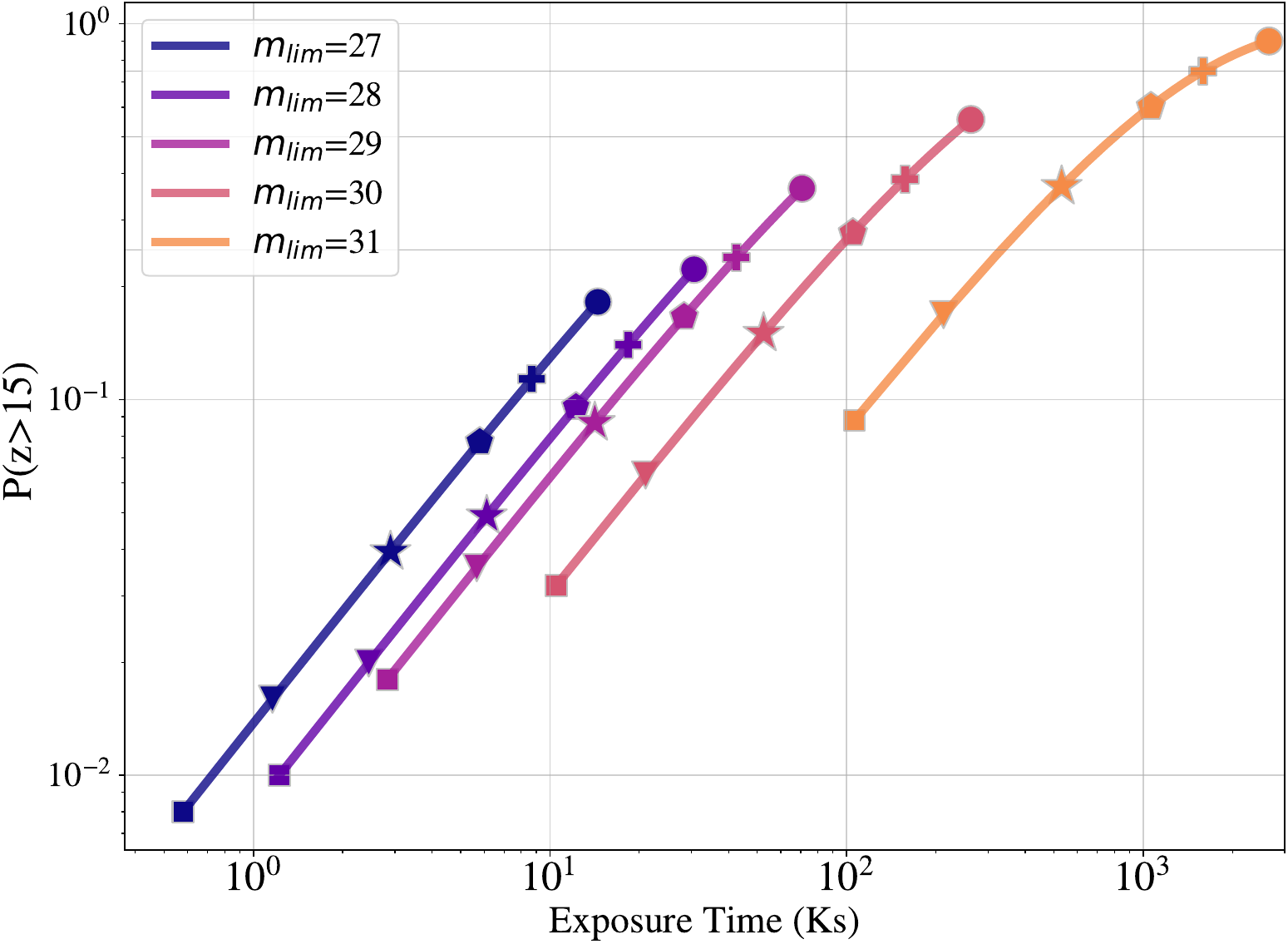}
        \caption{Probability of detecting a $z>15$ galaxy, $P(z>15)$, as a function of exposure time (ks) along the SMACS 0723 line of sight for a variety of survey depths. The markers represent the number of lines of sight required to achieve the given $P(z>15)$ as follows: the squares represent 1 line of sight, triangles represent 2, stars represent 5, pentagons represent 10, crosses represent 15, and circles represent 25 lines of sight. Although deep observations (requiring incredibly long exposure times) are necessary to approach $P(z>15) \sim 1$, one can achieve reasonable $P(z>15)$ by observing many cluster lenses at lower depths as opposed to fewer clusters at high depths.
        }
        \label{fig:obstime}
    \end{center}
\end{figure}

We see that there is a noticeable drop in efficiency at greater depths, indicating that it is more efficient to observe more clusters at shallower depths. However, at lower depths, even after observing along all 25 cluster lines of sight, the maximum $P(z>15)$ we can achieve is severely limited. Another major hindrance to a survey comprised of many shallow observations is that as the number of targets increases, the re-pointing time of the telescope increases as well, resulting in a longer total observing time.

 \subsection{Massive Galaxies as a Probe of $\Lambda$CDM}\label{subsec:probinglcdm}

Potentially the most impactful result of discovering the highest redshift galaxies will lie in their use as a cosmological probe. A series of recent studies have found hints of tension between the most massive high-redshift galaxies and the standard $\Lambda$CDM cosmological paradigm. The first stage in galaxy formation is the assembly of a compact halo, and the $\Lambda$CDM halo mass function is well constrained via numerical calculation \citep{2001MNRAS.323....1S, 2014Natur.509..177V}.  Subsequent processes turn the assembled baryons into stars and other structures that become recognizable as a galaxy.

However, the most massive sources at $z > 4$ from {\em Hubble} are so massive, so early, that there should not yet have been time for their halos to finish assembling, let alone to further turn their baryons into stars \citep{Steinhardt_2016}. This ``impossibly early" galaxy problem was sharpened by the discovery of high-redshift, massive quiescent galaxies \citep{2017Natur.544...71G}.  One proposed resolution was an increased stellar baryon fraction for the first galaxies \citep{2015ApJ...814...95F, Behroozi2018}. However, the tension increases towards high redshift, and initial {\em JWST} studies reported $z \sim 10$ galaxies so massive that even if all of their baryons had already ended up in stars, they would still be too massive to reconcile with the $\Lambda$CDM halo mass function \citep{Labb2023,BoylanKolchin2023}.  Although these redshifts and masses are not yet robustly proven \citep{steinhardt2022templates, 2023ApJ...943L...9Z}, and therefore may be easier to reconcile with $\Lambda$CDM \citep{casey2023,greene2023uncover, xiao2023massive, labe2023b, kokorev2023uncover}, it is clear that the most massive, earliest galaxies will provide a stringent test of the $\Lambda$CDM halo mass function and thus of $\Lambda$CDM. 


\subsection{The value of crossing the $z \sim 15$ threshold} \label{subsec:z15thres}

When pursuing the search for the most distant galaxy, it is important to consider the value in finding a $z > 15$ galaxy, which is possible in a cluster-lens field, as opposed to a $z \sim 14$ that could be found in a blank field. 
The range of redshifts between $15 \leq z \leq 20$ is incredibly important because our current cosmological models have placed the formation of the first galaxies in the Universe to be within this regime \citep{Springel_2005,Lacey_2011,Bromm_2011,Visbal_2012,2023arXiv230404348Y}. Thus, being able to find any galaxy in this redshift regime would be invaluable in constraining these models of early universe galaxy formation, as well as providing insight into the formation of the first generation of stars. Observing galaxies in this redshift regime could also lead to observational confirmation of Pop III stars, which are a hypothetical population of high-mass, low metallicity stars that are theorized to have formed around $20\leq z \leq 30$ \citep{Haiman_1997, Barkana_2001, Bromm2002, Visbal2018}. Observing these Pop III stars would be invaluable to understand the enrichment of the ISM in the early Universe, which led to the creation of the significantly more abundant Pop II stars which we can observe today. However, verifying the end of the cosmic dark ages would need to be confirmed with multiple sight-lines in order to get a robust limit.

\section*{Acknowledgements}
    
    The authors would like to thank Ana Acebron Mu\~noz, Mike Boylan-Kolchin, Gabriel Brammer, Jose Maria Diego, Vasily Kokorev, Bahram Mobasher, Anna Niemiec, Amanda Pagul, Vadim Rusakov, Claudia Scarlata, John Weaver, Liliya Williams and Radoslaw Wojtak for their helpful comments and discussions. This work is supported by VILLUM FONDEN grant no. 53101. The Cosmic Dawn Center (DAWN) is funded by the Danish National Research Foundation under grant No. 140.

    %
    %





\clearpage
\bibliography{bibliography}{}
\bibliographystyle{aasjournal}



\end{document}